\documentclass[10pt, aps, prd, amsmath, floats, floatfix, twocolumn, notitlepage,
superscriptaddress, nofootinbib, showpacs,groupedaddress]{revtex4-1}
\allowdisplaybreaks
\usepackage{float}
\usepackage[utf8]{inputenc}
\usepackage[T1]{fontenc}
\usepackage{comment}
\usepackage{bm}
\usepackage[normalem]{ulem}\usepackage{mathtools,amsmath,amssymb,amsfonts,mathrsfs,eucal,graphicx,tensor,csquotes,accents,commath,chngcntr,siunitx}
\usepackage[dvipsnames]{xcolor}
\usepackage[unicode]{hyperref}
\usepackage{orcidlink}
\hypersetup{colorlinks=true, citecolor=teal,
            linkcolor=teal, urlcolor=teal, linktocpage=true}
\usepackage[normalem]{ulem}

\DeclareMathAlphabet{\mathpzc}{OT1}{pzc}{m}{it}

\definecolor{darkgreen}{rgb}{0.0, 0.6, 0.0}

\newcommand{\iitgn}{Indian Institute of Technology Gandhinagar, Palaj 382055, India}

\labelformat{section}{Section #1} 
\labelformat{subsection}{Section \thesection.#1} 
\labelformat{subsubsection}{Section #1}
\labelformat{subsubsubsection}{Section #1}
\labelformat{equation}{Eq.~(#1)} 
\labelformat{figure}{Fig.~#1} 
\labelformat{subfigure}{Fig.~\thefigure#1} 
\labelformat{table}{Table~#1} 
\labelformat{appendix}{Appendix #1}
\labelformat{footnote}{Footnote #1} 

\begin{document}

\date{\today}

\title{Inverse Gertsenshteǐn-Zel'dovich Effect in Time-Independent Electromagnetic Field}

\author{Rajendra Prasad Bhatt\orcidlink{0009-0004-9088-2998}}
\email{rajendra@iitgn.ac.in}
\author{Rudraksh Verma\orcidlink{0009-0000-3439-0463}}
\email{rudraksh.verma@iitgn.ac.in}
\author{R. Prasad\orcidlink{0000-0002-6602-3913}}
\email{prasad@iitgn.ac.in}
\author{Sudipta Sarkar\orcidlink{0000-0002-4077-1358}}
\email{sudiptas@iitgn.ac.in}

\affiliation{\iitgn}

\begin{abstract}
The inverse Gertsenshteǐn-Zel'dovich effect describes the conversion of gravitational waves into electromagnetic waves in the presence of a background electromagnetic field. This effect has primarily been studied for constant background electromagnetic fields. In this work, we develop a general formalism for calculating the components of the electric and magnetic fields induced by the passage of a gravitational wave through a spatially varying, time-independent background electromagnetic field. The formalism accommodates all six possible polarization modes of gravitational waves. We illustrate its applicability through several examples, including gravitational waves propagating through constant and spatially varying electromagnetic fields. We further investigate the electromagnetic signals generated by the passage of gravitational waves through astrophysical magnetic fields, particularly those associated with neutron stars and magnetars. By connecting gravitational waves and electromagnetic waves--two essential messengers of multi-messenger astronomy--the inverse Gertsenshteǐn-Zel'dovich effect offers a promising framework for both gravitational-wave detection and astrophysical observations. The formalism developed here may therefore find applications in the design of high-frequency gravitational-wave detectors and in the investigation of high-energy astrophysical electromagnetic signals.
\end{abstract}

\maketitle

\section{Introduction}

In 1960, M. E. Gertsenshteǐn showed that electromagnetic waves (EMWs) can be converted into gravitational waves (GWs) in the presence of a magnetic field~\cite{Gertsenshtein:1962}. In the same work, he also mentioned the inverse conversion of GWs into EMWs, but regarded it as a problem ``hardly of interest''. In the following years, however, several authors investigated the inverse conversion. These include work by Lupanov, where he analysed the conversion of GWs to EMWs in the presence of an electric field~\cite{Lupanov:1967}, and the work by Boccaletti {\em et al.}, in which they studied the conversion of EMWs to GWs and vice versa in the presence of a static electromagnetic field~\cite{Boccaletti:1970pxw, Boccaletti:1971}. Zel'dovich further analysed this conversion and discussed the role of matter and coherence length~\cite{Zeldovich:1974} (see also~\cite{Macedo:1983wcr, Kolosnitsyn:2015zua, Ejlli:2020fpt}). Later, De Logi and Mickelson employed Feynman perturbation techniques to analyse the (inverse) GZ effect~\cite{DeLogi:1977qe}. In honour of Gertsenshteǐn and Zel'dovich, the conversion of EMWs to GWs is known as the Gertsenshteǐn-Zel'dovich (GZ) effect. In this work, we study the inverse conversion (GW$\to$EMW), known as the inverse GZ effect. Most existing studies of the inverse GZ effect have considered constant background electromagnetic fields. Here, we investigate this effect in a general time-independent, spatially-varying electromagnetic field, considering all six possible polarization modes of GWs.

The conversion of EMWs to GWs, and vice versa, can be qualitatively understood as follows: when an EMW (with field strength $F$) propagates through a background electromagnetic field $F^{(0)}$, it contributes a term proportional to $FF^{(0)}$ to the stress-energy tensor, which acts as a source for the induced GW. Similarly, in the inverse conversion, when a GW ($h$) propagates through a background electromagnetic field $F^{(0)}$, it produces stretching and compression of the electromagnetic field. This gives rise to a term proportional to $hF^{(0)}$, which acts as a source for the induced EMW~\cite{Zeldovich:1974}.

The detection of gravitational waves from compact binary coalescences by the LIGO-Virgo-KAGRA (LVK) collaboration has opened a new window for {\em observing} the universe~\cite{LIGOScientific:2026sit}. Like EMWs, GWs span a broad range of frequencies. While the LVK detectors are sensitive to a particular frequency range, approximately $\mathcal{O}(10-1000)~{\rm Hz}$, several other detectors have been proposed or commissioned to explore other regions of the GW spectrum~\cite{LISA:2017pwj,Baker:2019nia,Kawamura:2018esd,InternationalPulsarTimingArray:2023mzf,Vacalis:2023gdz,Bringmann:2023gba,Capdevilla:2025omb, Wenskat:2026zjq}. At the low-frequency end, evidence for a stochastic GW background has been reported in Pulsar Timing Array (PTA) data~\cite{InternationalPulsarTimingArray:2023mzf}. Moreover, Goryachev {\em et al.} reported two events in a bulk acoustic wave antenna and interpreted them as possible High Frequency GW (HFGW) signals~\cite{Goryachev:2021zzn} (see also Refs.~\cite{Lasky:2021naa, Domenech:2021odz, Campbell:2025mks} for further discussions and debates concerning these detections). However, an unambiguous detection of GWs outside the frequency band of ground-based LVK detectors remains elusive. Since the inverse GZ effect provides a mechanism for converting GWs into EMWs in the presence of a background electromagnetic field, it has been explored as a possible basis for GW detection. Such proposals date back to the work by Lupanov, who considered the use of capacitors~\cite{Lupanov:1967}. Since the Poynting flux of the EMW induced through the inverse GZ effect is proportional to the square of the frequency of the incident GW, detectors based on this effect have primarily been proposed for detecting HFGWs~\cite{Cruise:2012zz,Zheng:2017fkv, Li:2017jcz, Li:2023tzw, Vacalis:2023gdz, Cai:2025fpe}.

The inverse GZ effect may also have astrophysical implications. The universe contains various sources of electromagnetic fields, such as stellar magnetospheres and large-scale astrophysical magnetic fields, as well as sources of GWs, including compact binary coalescences. Consequently, the propagation of GWs through such electromagnetic fields can, in principle, generate observable electromagnetic radiation. Several authors have investigated this possibility. In Ref.~\cite{Marklund:1999sp}, the interaction of weak GWs with electromagnetic fields in a thin plasma was studied, and the resulting radiation was found to lie in the radio-frequency band. In Ref.~\cite{Dolgov:2012be}, the probability of graviton-photon conversion in cosmological background magnetic fields during recombination and subsequent cosmological epochs was investigated, with the resulting electromagnetic radiation found to be concentrated in the X-ray region of the spectrum. In Ref.~\cite{Kushwaha:2022twx}, the inverse GZ effect was proposed as a plausible explanation for a class of non-repeating fast radio bursts (FRBs) (see also Refs.~\cite{Kalita:2022uyu,Wu:2026nds}).

Moreover, the inverse GZ effect has also been employed to constrain (ultra-) HFGWs~\cite{Ejlli:2019bqj,Ito:2023fcr,Ito:2023nkq,Dandoy:2024oqg,Domcke:2020yzq,He:2023xoh,McDonald:2024nxj,Domcke:2025qlw,Kushwaha:2025mia}. In Refs. \cite{Ejlli:2019bqj,Ito:2023fcr,Ito:2023nkq,Dandoy:2024oqg}, upper limits on the stochastic GW background were obtained. In Ref.~\cite{Domcke:2020yzq}, it was shown that the conversion of GWs to EMWs can lead to a distortion of the cosmic microwave background (CMB), providing a possible probe of HFGW sources active before reionization. Constraints on the GW amplitude were obtained using observations from the radio experiments EDGES~\cite{Bowman:2018yin} and ARCADE 2~\cite{Fixsen:2011}. In Ref.~\cite{He:2023xoh}, upper limits on GW energy densities were derived from astrophysical observations using the inverse GZ effect. The effect has also been used to constrain the circular polarization of HFGWs through CMB observations~\cite{Kushwaha:2025mia}. In Ref.~\cite{McDonald:2024nxj}, the resonant conversion of GWs into photons in the neutron star magnetosphere was investigated, and observational limits were placed on the amplitude of a possible stochastic GW background (see also Ref.~\cite{Domcke:2025qlw}). Furthermore, the inverse GZ effect has been studied in the context of black hole ringdown~\cite{Clarkson:2003mp}. In Ref.~\cite{Ejlli:2018hke}, graviton-photon mixing was investigated in cosmological magnetic fields in an expanding universe. In addition, Ref.~\cite{Cembranos:2023ere} examined the (inverse) GZ effect in a background cosmological magnetic field within different theories of gravity. 

Thus, the inverse GZ effect has a broad range of applications, extending from the development of HFGW detectors to the investigation of astrophysical EM signals. An important feature of this effect is that it connects two major observational channels of multi-messenger astronomy: GWs and EMWs. It may therefore provide a useful framework for investigating fundamental gravitational physics. Furthermore, with the development of high-frequency single photon detectors~\cite{Komiyama:2011}, the inverse GZ effect could potentially play a role in proposals for single graviton detection, offering possible experimental insights into the quantization of gravity~\cite{Tobar:2023ksi}.

Most existing studies of the inverse GZ effect focused on constant background electromagnetic fields. In this work, we develop a formalism for calculating the induced electric and magnetic fields generated when a GW propagates through a general time-independent, spatially varying background electromagnetic field (\ref{Section: Formalism}). Our formalism incorporates all six GW polarization modes: plus, cross, breathing, longitudinal, vector$-x$, and vector$-y$. It also applies to GWs within Einstein's theory of General Relativity (GR) with only the two tensor polarization modes, plus and cross, by setting the contributions from the additional modes to zero. We also illustrate the formalism through several applications (\ref{Section: Applications}), including GW propagation through a constant background electromagnetic field (\ref{Section: Constant background electric and magnetic field}) and a time-independent magnetic field with linear spatial dependence (\ref{Section: Time-independent magnetic field}). Finally, in \ref{Section: GW passing through astrophysical magnetic fields: neutron star magnetosphere}, we investigate the astrophysical electromagnetic signals generated by the inverse GZ effect in a stellar magnetosphere, including their polarization properties. We conclude with a discussion of our results and possible directions for future work (\ref{Section: Discussion}). Some expressions used in the main text are presented in the appendices.

{\em Notations and Convention:} In this work, we have used the positive signature convention, i.e., the Minkowski metric in the $3+1$-dimensional spacetime is given by diag.$(-1,+1,+1,+1)$. We work in C.G.S. units. Vector quantities are shown in boldface. Moreover, $\mathbb{N}_0$ denotes the set of natural numbers including zero: $0,1,2,\dots$. Various terms in the paper appear with numbered subscripts, e.g., $C_{1},~C_{2}$. In a few places, we have denoted them collectively by $C_{1,2}$. Similarly, $\widetilde{e/b}_{x,y,z}(x,y,z)$ denotes the spatial $(x,y,z)$ components of $\bm{\widetilde{e}}(x,y,z)$ and $\bm{\widetilde{b}}(x,y,z)$.

\section{Formalism}\label{Section: Formalism}
Let us consider a gravitational wave (GW) propagating along the positive $z$ direction. In the linearized approximation, the metric tensor takes the form
\begin{equation}
    g_{\mu\nu} = \eta_{\mu\nu} + h_{\mu\nu}, \qquad g^{\mu\nu} = \eta^{\mu\nu} - h^{\mu\nu}; \qquad h\ll 1.
\end{equation}
Here, $\eta_{\mu\nu}$ denotes the Minkowski metric, $\eta_{\mu\nu} = \eta^{\mu\nu} = \rm{diag.}(-1,+1,+1,+1)$. The metric perturbation $h_{\mu\nu}$ can be expressed in terms of the possible GW polarizations as~\cite{Will:2014kxa,Will_2018,Isi:2022mbx} 
\begin{equation}\label{h}
    h_{\mu\nu} = h^{\mu\nu} \equiv \begin{bmatrix}
0 & 0 & 0 & 0\\
0 & h_b + h_+ & h_\times & h_x\\
0 & h_\times & h_b - h_+ & h_y\\
0 & h_x & h_y & h_l
\end{bmatrix}.
\end{equation}
Each $h_j;~j = \{+,\times,b,l,x,y\}$ is a function of $(t,z)$, where $z$ denotes the GW propagation direction. We consider the general form
\begin{equation}\label{form_of_hj}
    h_j = A_j e^{i(k_g z -\omega_g t)}, \qquad j = +,\times,b,l,x,y.
\end{equation}
The $+$ and $\times$ modes are the standard plus and cross modes of polarization, whereas $b$, $l$, $x$, and $y$ represent the breathing, longitudinal, vector-$x$, and vector-$y$ polarization modes, respectively. These additional modes may arise in modified theories of gravity~\cite{Eardley:1973br,Eardley:1973zuo,Nishizawa:2009bf,Will:2014kxa,Will_2018,Isi:2022mbx}. Importantly, in \ref{form_of_hj} we do not impose $\omega_g = k_g c$. This allows the GW propagation speed to differ from the speed of light, while retaining $\omega_g = k_g c$ as an important special case. We discuss the motivation for this choice later.

We now introduce time-independent background electric and magnetic fields, denoted by $\bm{E^{(0)}}$ and $\bm{B^{(0)}}$, respectively. The corresponding fields induced through the inverse GZ effect are denoted by $\bm{\widetilde{E}}$ and $\bm{\widetilde{B}}$. The total electromagnetic field tensor is:
\begin{equation}\label{EMT_P}
    F_{\mu\nu} = \begin{bmatrix}
0 & -E_x & -E_y & -E_z\\
E_x & 0 & B_z & -B_y\\
E_y & -B_z & 0 & B_x\\
E_z & B_y & -B_x & 0
\end{bmatrix},
\end{equation}
where $F_{\mu\nu} = F_{\mu\nu}^{(0)} + \widetilde{F}_{\mu\nu}$, and $X_i= X_i^{(0)}+\widetilde{X}_i$, for $X=E,B$; the superscript $(0)$ labels the background contribution. The background fields are assumed to be time-independent, with spatial dependence on $x,y,z$ only. Since the time dependence comes only from the exponential terms in the incident GW, we use the same time dependence for the induced electromagnetic fields and write
\begin{equation}\label{Separation}
    \widetilde{E/B}_{x,y,z}(t,x,y,z) = \widetilde{e/b}_{x,y,z}(x,y,z) e^{i(k_g z -\omega_g t)}.
\end{equation}

From now on we will remove the argument $(x,y,z)$ from $\widetilde{e/b}_{x,y,z}$ for notational simplicity. Our aim is to solve Maxwell's equations for these field amplitudes.

The source-free Maxwell equations in curved spacetime read\footnote{Real astrophysical systems, such as neutron star magnetospheres, may contain plasma, which would require the inclusion of source terms. Our formalism applies to the source-free case and may be useful in certain limits, such as in low-density plasma.}
\begin{equation}\label{SF_ME}
    \partial_\mu(\sqrt{-g}\,F^{\mu\nu})=0, \qquad \partial_\mu(\sqrt{-g}\,G^{\mu\nu}) = 0,
\end{equation}
where $G^{\mu\nu}$ is the dual electromagnetic field tensor, defined by $2\sqrt{-g}\,G^{\mu\nu} = \eta^{\mu\nu\alpha\beta}F_{\alpha\beta}$, with $\eta^{\mu\nu\alpha\beta}$ being the antisymmetric Levi-Civita symbol, normalized by $\eta^{0123} = -\eta_{0123} = 1$.

Substituting the metric perturbation and the electromagnetic field decomposition into the first Maxwell equation gives 
\begin{equation}\label{Eq_8}
    \partial_x \widetilde{e}_x+\partial_y \widetilde{e}_y +(\partial_z + ik_g)(\widetilde{e}_z + \mathcal{S}_E)
    - \partial_x \,_{E}\mathcal{S}^{+}_{1}
    +\partial_y \,_{E}\mathcal{S}^{-}_{2} = 0
\end{equation}
\begin{equation}
    \partial_y \widetilde{b}_z -\partial_z \widetilde{b}_y +i k_g (\gamma\widetilde{e}_x - \widetilde{b}_y) 
    +\partial_z \,_{B}\mathcal{S}^{-}_{2}
    -\partial_y \mathcal{S}_B
    + 2\mathcal{R}_+^{(2)} = 0
\end{equation}
\begin{equation}
    \partial_z \widetilde{b}_x- \partial_x \widetilde{b}_z + i k_g (\gamma\widetilde{e}_y + \widetilde{b}_x) 
    +\partial_z \,_{B}\mathcal{S}^{+}_{1}
    + \partial_x\,\mathcal{S}_B 
    - 2\mathcal{R}_+^{(1)}= 0
\end{equation}
\begin{equation}\label{Eq_10}
    \partial_x \widetilde{b}_y-\partial_y \widetilde{b}_x+i \frac{\omega_g}{c} \widetilde{e}_z-\partial_x \,_{B}\mathcal{S}^{-}_{2}
    -\partial_y \,_{B}\mathcal{S}^{+}_{1}
    +i\frac{\omega_g}{c}\mathcal{S}_E = 0,
\end{equation}
where $\gamma = \omega_g/k_g c$.
For compactness, we introduce the spatially dependent source quantities $\mathcal{S}_{E,B}$, $\,_{E,B}\mathcal{S}^{\pm}_{1,2}$, $\mathcal{R}_\pm^{(1,2)}$, defined by
\begin{equation}\label{Eq_11}
    \mathcal{S}_X = A_b X_z^{(0)}- \frac{A_l}{2} X_z^{(0)}-A_x X_x^{(0)} -A_y X_y^{(0)},
\end{equation}
\begin{multline}
    \,_{X}\mathcal{S}^{\pm}_{j} = A_+ X_j^{(0)} \pm A_\times X_{j\pm1}^{(0)} \mp \frac{A_l}{2} X_j^{(0)} + (2-j) A_x X_z^{(0)} 
    \\
    - (j-1) A_y X_z^{(0)},
\end{multline}
\begin{multline}
    \mathcal{R}_\pm^{(1)} = -\frac{ik_g}{2} \left\{A_+(\gamma E_y^{(0)}\pm B_x^{(0)}) - A_\times ( \gamma E_x^{(0)}\mp  B_y^{(0)}) 
    \right.\\\left.
    + \frac{A_l}{2}(\gamma E_y^{(0)} \mp B_x^{(0)}) \pm A_x B_z^{(0)} - \gamma A_y E_z^{(0)}\right\},
\end{multline}
and
\begin{equation}\label{Eq_14}
    \mathcal{R}_\pm^{(2)} = \mp\,\mathcal{R}_\mp^{(1)} (\gamma E^{(0)}\leftrightarrow B^{(0)}).
\end{equation}
In these definitions, $X= \{E,B\}$ and subscript $j = \{1,2,3\}$ corresponds to $\{x,y,z\}$. The quantities $\,_{X}\mathcal{S}^{-}_{1}$ and $\,_{X}\mathcal{S}^{+}_{2}$ do not appear in the Maxwell equations.

\ref{Eq_8} to \ref{Eq_10} form what we call set A. In these equations, we have neglected terms of $\mathcal{O}(h^2, \widetilde{F}^2, h\widetilde{F})$. Separating the time-dependent part from the original Maxwell equation also yields two Maxwell equations for the background time-independent electromagnetic field, viz.,
\begin{equation}\label{BM1}
    \bm{\nabla}\cdot\bm{E^{(0)}} = 0, \qquad \bm{\nabla}\times\bm{B^{(0)}} = 0.
\end{equation}

The second (dual) Maxwell equation similarly yields
\begin{equation}
    \partial_x \widetilde{b}_x+\partial_y \widetilde{b}_y+\partial_z \widetilde{b}_z + ik_g \widetilde{b}_z = 0,
\end{equation}
\begin{equation}
    \partial_y \widetilde{e}_z -\partial_z \widetilde{e}_y - ik_g (\gamma\widetilde{b}_x + \widetilde{e}_y) = 0,
\end{equation}
\begin{equation}
    \partial_z \widetilde{e}_x - \partial_x \widetilde{e}_z + ik_g (\widetilde{e}_x - \gamma\widetilde{b}_y) = 0,
\end{equation}
\begin{equation}
    \partial_x \widetilde{e}_y - \partial_y \widetilde{e}_x - i\frac{\omega_g}{c} \widetilde{b}_z  = 0.
\end{equation}
The four equations constitute set B, with terms of $\mathcal{O}(h^2, \widetilde{F}^2, h\widetilde{F})$ being neglected. Separating the time-dependent part from the original (dual) Maxwell equation yields the remaining Maxwell equations for the background time-independent electromagnetic field, viz.,
\begin{equation}\label{BM2}
    \bm{\nabla}\cdot\bm{B^{(0)}} = 0, \qquad \bm{\nabla}\times\bm{E^{(0)}} = 0.
\end{equation}
The fourth equations of Sets A and B determine the longitudinal field amplitudes, $\widetilde{e}_z$ and $\widetilde{b}_z$, in terms of the spatial derivatives of the transverse components $\widetilde{b}_x$, $\widetilde{b}_y$ and $\widetilde{e}_x$, $\widetilde{e}_y$, respectively. We therefore first solve for the transverse components, which will consequently yield the longitudinal components.

To do so, we substitute the expressions for $\widetilde{e}_z$ and $\widetilde{b}_z$ into the first three equations of both sets. The first equation of Set B can then be written as 
\begin{equation}\label{Eq_22}
    \widetilde{b}_x -\mathcal{D} \widetilde{e}_y= \partial_y \phi_1, \qquad \widetilde{b}_y +\mathcal{D}\widetilde{e}_x = - \partial_x \phi_1,
\end{equation}
where $\mathcal{D}$ denotes the differential operator
\begin{equation}
    \mathcal{D} = \frac{ic}{\omega_g} \left(\partial_z + ik_g\right).
\end{equation}
Likewise, the first equation of Set A gives
\begin{equation}\label{Eq_24}
    \widetilde{e}_x +\mathcal{D}\widetilde{b}_y -\mathcal{D}\,_{B}\mathcal{S}^{-}_{2}-\,_{E}\mathcal{S}^{+}_{1} = -\partial_y \phi_2
\end{equation}
and
\begin{equation}\label{Eq_25}
    \widetilde{e}_y -\mathcal{D}\widetilde{b}_x -\mathcal{D}\,_{B}\mathcal{S}^{+}_{1}
    +\,_{E}\mathcal{S}^{-}_{2} = \partial_x \phi_2,
\end{equation}
where $\phi_1$ and $\phi_2$ are unknown functions of the spatial coordinates $(x,y,z)$.~\footnote{While solving the differential equations, we encounter relations of the form $\partial_x \mathcal{G}_1(x,y,z) = \partial_y \mathcal{G}_2(x,y,z)$. We represent them through $\mathcal{G}_1(x,y,z) = \partial_y \phi(x,y,z)$ and $\mathcal{G}_2(x,y,z) = \partial_x \phi(x,y,z)$. This procedure introduces the unknown functions $\phi_{1,2}$, while $f_{1,2,3,4}(x,y)$ (see \ref{Appendix: Some other expressions}) arise as the "constants" of integration with respect to $z$ at intermediate stages. We return to this footnote later.\label{Footnote_2}}

Combining \ref{Eq_22}, \ref{Eq_24}, and \ref{Eq_25}, we obtain four first-order equations for the combinations of transverse field amplitudes:
\begin{equation}
    (\mathcal{D}-1)(\widetilde{e}_y+\widetilde{b}_x) = -\partial_y \phi_1 - \partial_x \phi_2 - \mathcal{D}\,_{B}\mathcal{S}^{+}_{1}
    +\,_{E}\mathcal{S}^{-}_{2},
\end{equation}
\begin{equation}
    (\mathcal{D}+1)(-\widetilde{e}_y+\widetilde{b}_x) = \partial_y \phi_1 - \partial_x \phi_2 - \mathcal{D}\,_{B}\mathcal{S}^{+}_{1}
    +\,_{E}\mathcal{S}^{-}_{2},
\end{equation}
\begin{equation}
    (\mathcal{D}+1)(\widetilde{e}_x +\widetilde{b}_y) = - \partial_x \phi_1 -\partial_y \phi_2 + \mathcal{D}\,_{B}\mathcal{S}^{-}_{2} + \,_{E}\mathcal{S}^{+}_{1},
\end{equation}
and
\begin{equation}\label{Eq_28}
    (\mathcal{D}-1)(\widetilde{e}_x -\widetilde{b}_y) = - \partial_x \phi_1 +\partial_y \phi_2 - \mathcal{D}\,_{B}\mathcal{S}^{-}_{2} - \,_{E}\mathcal{S}^{+}_{1}.
\end{equation}

These first-order equations can be solved to obtain expressions for $\widetilde{e/b}_{x,y}$. The longitudinal components $\widetilde{e/b}_{z}$ then follow from the fourth equations of sets A and B.
\footnote{When $\omega_g = k_g c$, one may alternatively proceed by expressing $\widetilde{b}_x + \widetilde{e}_y$ and $\widetilde{b}_y - \widetilde{e}_x$ using the first equation of set B, in terms of $\partial_z \widetilde{e}_y$ and $\partial_z \widetilde{e}_x$, respectively. The first equation of set A expresses the same combination through $\partial_z \widetilde{b}_x$ and $\partial_z \widetilde{b}_y$, respectively. Combining these relations determines $\widetilde{e}_x + \widetilde{b}_y$ and $\widetilde{e}_y - \widetilde{b}_x$ and hence $\widetilde{e/b}_{x,y,z}$ for $\omega_g = k_g c$.} These resulting expressions are provided in \ref{Appendix: Some other expressions}. 

The expressions of $\widetilde{e/b}_{x,y,z}$ still contain some unknown functions $\phi_1(x,y,z)$, $\phi_2(x,y,z)$, $f_1(x,y)$, $f_2(x,y)$, $f_3(x,y)$, and $f_4(x,y)$. To constrain them, we now impose the remaining Maxwell equations: the second and third equations of both sets.

The field components $\widetilde{e/b}_{x,y,z}$ (given in \ref{Appendix: Some other expressions}) satisfy the first and last equations of sets A and B. Substituting them into the second and third equations of set A yields
\begin{equation}\label{Condition_1}
    \partial_y \widetilde{\mathcal{Q}}_1(x,y,z) = 0, \qquad {\rm and} -\partial_x \widetilde{\mathcal{Q}}_1(x,y,z) = 0,
\end{equation}
respectively. Similarly, the second and third equations of set B yield
\begin{equation}\label{Condition_2}
    \partial_y \widetilde{\mathcal{Q}}_2(x,y,z) = 0, \qquad {\rm and} -\partial_x \widetilde{\mathcal{Q}}_2(x,y,z) = 0,
\end{equation}
respectively. Explicit expressions for $\widetilde{\mathcal{Q}}_1(x,y,z)$ and $\widetilde{\mathcal{Q}}_2(x,y,z)$ are provided in \ref{Appendix: Some other expressions}. 

These conditions require $\widetilde{\mathcal{Q}}_1(x,y,z)$ and $\widetilde{\mathcal{Q}}_2(x,y,z)$ to depend only on $z$ (the direction of GW propagation). However, the expressions for $\widetilde{\mathcal{Q}}_1(x,y,z)$ and $\widetilde{\mathcal{Q}}_2(x,y,z)$ involve functions of $x$, $y$, and $z$. Thus, \ref{Condition_1} and \ref{Condition_2} impose additional constraints that determine some of the unknown functions.

Although the expressions of $\widetilde{\mathcal{Q}}_1(x,y,z)$ and $\widetilde{\mathcal{Q}}_2(x,y,z)$ are lengthy in general, they can simplify substantially for particular background electromagnetic fields. We demonstrate these simplifications in \ref{Section: Applications}. Once these constraints are satisfied, the resulting fields obey all the Maxwell equations. Therefore, they provide a general representation of the electromagnetic fields induced by the inverse GZ effect when a GW with all possible polarizations propagates through a time-independent, spatially varying background electromagnetic field.

The result also includes the general relativistic case, in which only the plus and cross polarization modes are present and they propagate at the speed of light. To obtain the expressions for general relativity, we need to set the terms related to the extra polarization modes to zero, i.e., $A_{b,l,x,y} =0$, and impose $\omega_g = k_g c$, which means that the GW travels at the speed of light. 

At the outset, we highlight a noteworthy property of this solution for the usual plus and cross GW polarization modes: the induced electric and magnetic fields depend only on the transverse $x,y$ components of the background electric and magnetic field, without involving the spatial derivatives. When additional GW polarizations are included, the $z$ components of the background electromagnetic field also enter through the extra polarizations. These additional contributions are absent when only the two tensor polarizations (plus and cross) are present, as expected from the classic inverse GZ effect, which requires a background electric or magnetic field perpendicular to the GW propagation direction. Thus, {\em if a background electric or magnetic field is present only along the GW propagation direction, it gives rise to an induced EMW in the presence of suitable additional GW polarization modes.} We illustrate this possibility explicitly in \ref{Section: Constant background electric and magnetic field}. 

This section set up the formalism for the inverse GZ effect in a time-independent, spatially varying background electromagnetic field. In the subsequent section, we will discuss the applicability of the formalism in a few other interesting cases.

\subsection{Effects of the propagation speed of GW}\label{Section: Effects of the speed of GW propagation}
In classical general relativity (GR), GWs propagate at the speed of light, and the graviton is massless. However, the graviton can be massive in modified theories of gravity. In that case, GWs propagate at a speed lower than that of light~\cite{Visser:1997hd,Nishizawa:2009bf}:
\begin{equation}
    \frac{v_g(\omega)}{c} = \sqrt{1-\frac{m_g^2\,c^4}{\hbar^2\omega^2}},
\end{equation}
where $v_g(\omega)$ is the group velocity of GW, $m_g$ is the mass of the graviton, and $\hbar$ is the reduced Planck constant.

The graviton mass is constrained by several observations~\cite{Finn:2001qi, Bernus:2019rgl, LIGOScientific:2019fpa, LIGOScientific:2020tif, LIGOScientific:2021sio, LIGOScientific:2026fcf} (see Ref.~\cite{deRham:2016nuf} for a review). Recent results from the LVK collaboration constrain the graviton mass to be below $1.92\times10^{-23}~{\rm eV/c^2}$ at a $90\%$ credibility level~\cite{LIGOScientific:2026fcf}, which yields $v_g(\omega) = c\sqrt{1-8.51\times 10^{-16}\,\omega^{-2}}$. Hence, a non-zero graviton mass will cause the speed of GWs to differ only slightly from the speed of light. Since our formalism does not assume that the GW propagates at the speed of light, it will also hold for GWs propagating at other speeds. Also in GR, GWs have only two polarization modes, $+$ and $\times$, and they propagate at the speed of light. However, modified theories of gravity allow for up to six polarization modes~\cite{Nishizawa:2009bf}, and these modes may propagate at different speeds~\cite{Schumacher:2023jxq}. These differences have not been constrained due to the lack of detection of the different modes. Based on the constraint on the GW propagation speed derived from the graviton mass bound, we expect the differences in propagation speeds among the different polarization modes to be quite small, i.e., they should all travel nearly at the speed of light. 

However, our proposed formalism in \ref{Section: Formalism} can also be applied in this setup, where different polarization modes propagate at different speeds. To compute the components of the induced electromagnetic fields, we apply the formalism separately to each mode. In this study, we consider one mode at a time (say, mode $j$), setting the parameters associated with all other modes to zero and replacing the wave number $k_g$ with the corresponding wave number $k_j$ for that mode in the solution obtained in the previous section. This procedure yields the field amplitudes induced by that mode, $\widetilde{e/b}_{x,y,z}^{(j)}$. We then repeat the same procedure for the remaining modes to obtain their corresponding field amplitudes. It is important to note that each mode has its own set of unknown functions, $\phi_{1,2}^{(j)}(x,y,z)$ and $f_{1,2,3,4}^{(j)}(x,y)$. Therefore, \ref{Condition_1} and \ref{Condition_2} must be satisfied separately for each mode.

A general solution can be obtained by summing the individual components of the induced electromagnetic fields from all polarization modes. If $\bm{\widetilde{E}^{(j)}}$ and $\bm{\widetilde{B}^{(j)}}$ denote the induced electric and magnetic fields, respectively, due to the $j^{\rm th}$ mode (where $j=+,\times,b,l,x,y$), then the general solution can be written as
\begin{equation}
    \widetilde{E/B}_{n} = \sum_{j=+,\times,b,l,x,y} \widetilde{E/B}_{n}^{(j)}, \qquad n=x,y,z.
\end{equation}
This will provide the resultant electromagnetic field induced by a GW propagating through a spatially dependent, time-independent electromagnetic field, with all six possible GW polarization modes propagating at different speeds.

We will now illustrate the formalism presented in \ref{Section: Formalism} by applying it to several examples to calculate the induced EMWs. In these applications, we assume that the GW propagates at the speed of light. We will discuss the effects of differences in propagation speeds among the polarization modes in \ref{Section: Discussion}.

\section{Applications}\label{Section: Applications}
In this section, we will illustrate the formalism presented in this paper through several applications. In these applications, we assume that all GW polarization modes propagate at the speed of light, i.e.,
\begin{equation}
    \omega_g = k_g c \,.
\end{equation}
This is just to keep the calculations simple, and as we discussed in the formalism section, we can immediately generalize all the calculations for $ \omega_g \neq k_g c$. First, we will discuss a classic example of a GW passing through a constant background electromagnetic field. Next, we will consider a different configuration of the background electromagnetic field, in which specific components of the magnetic field depend linearly on the spatial coordinates. Finally, we will discuss the astrophysical EM signals associated with a GW passing through an astrophysical magnetic field. The titles of the following subsections describe the magnetic fields through which the GWs will propagate.

\subsection{Constant background electric and magnetic field}\label{Section: Constant background electric and magnetic field}
Let us consider a GW propagating through constant background electric and magnetic fields given by
\begin{equation}
    E_{x,y,z}^{(0)}(x,y,z) = E_{x,y,z}^{(0)}, \,\,\,\,\,\,\,\, B_{x,y,z}^{(0)}(x,y,z) = B_{x,y,z}^{(0)}.
\end{equation}
Since the background electric and magnetic fields are constant, the spatial dependence in the solution appears only through $ z$, owing to the GW propagation direction. Hence, we can replace each unknown function of $x,y$ appearing in the expressions for  $\widetilde{e/b}_{x,y,z}$ (see \ref{Appendix: Some other expressions}) with a constant: $f_{1,2,3,4}(x,y) = f_{1,2,3,4}$. 

There are a few other unknown functions of $x,y,z$, namely, $\phi_{1,2}$, appearing in the expressions for $\widetilde{e/b}_{x,y,z}$. If we assume that $\phi_{1,2}$ depend only on $z$, we might miss some functions of $z$ in our analysis (see \ref{Footnote_2}). Thus, we write
\begin{equation}
    \phi_j(x,y,z) = \phi_{j0}(z) + x\,\phi_{j1}(z) + y\,\phi_{j2}(z);\, j = 1,2.
\end{equation}

Now imposing \ref{Condition_1} and \ref{Condition_2} yields 
\begin{equation}
    \phi_{11}(z) = \phi_{12}(z) = \phi_{21}(z) = \phi_{22}(z) = 0.
\end{equation}
This satisfies all the constraint equations, and our resulting solution obeys the Maxwell equations. Having obtained the forms of a few unknown functions, the solution is
\begin{equation}\label{Eq_31}
    \widetilde{e}_x = \frac{f_4}{2} e^{-2 i k_g z}+z \mathcal{R}_+^{(2)}
    +\frac{i}{2k_g} \mathcal{R}_-^{(2)} - \frac{\,_B\mathcal{S}_2^-}{2}+ \frac{f_3}{2},
\end{equation}
\begin{equation}
    \widetilde{e}_y = \frac{f_1}{2} e^{-2 i k_g z}-z \mathcal{R}_+^{(1)} -\frac{i}{2k_g}\mathcal{R}_-^{(1)}- \frac{\,_B\mathcal{S}_1^+}{2}- \frac{f_2}{2},
\end{equation}
\begin{equation}
    \widetilde{b}_x = \frac{f_1}{2} e^{-2 i k_g z}+ z \mathcal{R}_+^{(1)}
    -\frac{i}{2k_g}\mathcal{R}_+^{(1)} + \frac{f_2}{2},
\end{equation}
\begin{equation}
    \widetilde{b}_y = - \frac{f_4}{2} e^{-2 i k_g z}+ z \mathcal{R}_+^{(2)}
    -\frac{i\mathcal{R}_+^{(2)}}{2k_g} +\frac{f_3}{2},
\end{equation}
\begin{equation}\label{Eq_36}
    \widetilde{e}_z = -\mathcal{S}_E, \qquad \widetilde{b}_z = 0,
\end{equation}
where the source terms are defined in \ref{Eq_11}$-$\ref{Eq_14}. It is important to note that, although our ansatz (\ref{Separation}) appears to select an induced wave propagating along the positive $z$-direction, it does not exclude propagation in the opposite direction. The field amplitudes $\widetilde{e/b}_{x,y,z}$ depend on $x,y,z$ and give rise to terms proportional to ${\rm exp}(-2ik_g z)$, thereby allowing an induced EMW propagating along the negative $z$-direction, as can be seen from the solutions above.

\begin{figure}[h!]
    \centering
    \includegraphics[scale=0.5]{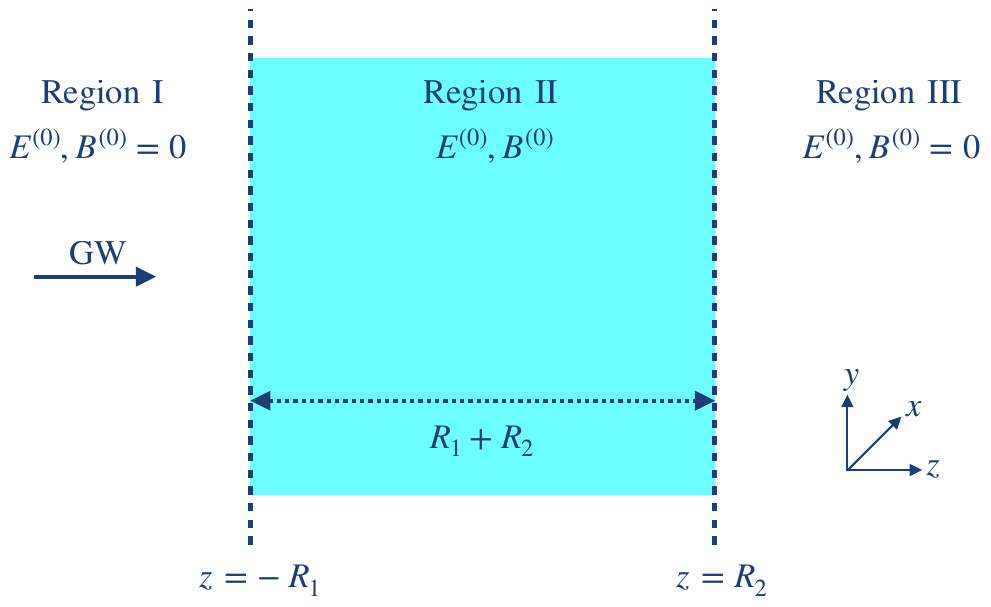}
    \caption{Schematic illustration of the different regions: the background electromagnetic field is non-zero in region II and zero in regions I and III. The GW propagates along the positive $z$-direction, entering from region I. The coordinates $z = -R_1$ and $R_2$ specify the locations of the boundaries of region II, where $R_{1,2}\ge 0$. Hence, the width of region II is $R_1+R_2$. The coordinate system is also shown in the bottom right corner, where the $y$- and $z$-axes lie in the plane of the paper, while the $x$-axis is perpendicular to the plane and points inward.}
    \label{Figure:Different regions}
\end{figure}

Note that there are a few unknown constants that need to be fixed by matching at the boundaries. Let us assume that the background electromagnetic field is confined to the region $-R_1\le z \le R_2$, where $R_{1,2}\ge 0$. We can divide the entire space into three regions (see \ref{Figure:Different regions}):
\begin{itemize}
    \item Region I: $z\le -R_1$.
    \item Region II: $-R_1\le z \le R_2$.
    \item Region III: $z\ge R_2$.
\end{itemize}

Since there are no background electric or magnetic fields in regions I and III, the solution in each of these regions will be a plane wave. In region I, the wave will propagate in the negative $z$-direction, whereas in region III, it will propagate in the positive $z$-direction. 

The solution in region I is
\begin{equation}
    \widetilde{e}_x = -\widetilde{b}_y = C_1 e^{-2ik_g z},
\end{equation}
and
\begin{equation}
    \widetilde{e}_y = \widetilde{b}_x = \widetilde{C}_1 e^{-2ik_g z},\qquad \widetilde{e}_z = \widetilde{b}_z = 0.
\end{equation}

Similarly, the solution in region III is
\begin{equation}
    \widetilde{e}_x = \widetilde{b}_y = C_3, \,\,\,\,\,\,\, \widetilde{e}_y = -\widetilde{b}_x = \widetilde{C}_3, \,\,\,\,\,\,\, \widetilde{e}_z = \widetilde{b}_z = 0,
\end{equation}
where $C_{1,3}$ and $\widetilde{C}_{1,3}$ are unknown constants. Moreover, \ref{Eq_31}-\ref{Eq_36} give the solution in region II. 

Since we are working in curved spacetime (in the linearized gravity regime), the quantities we should match at the boundaries are $-\sqrt{-g}\,F^{j0}$ and $-\sqrt{-g}\,G^{j0}$, where $j=1,2,3$. Following the notation introduced in \ref{Section: Formalism}, these quantities are \footnote{Note that for $j=1,2,3$, $-\sqrt{-g}\,F^{j0}$ becomes $(Q_j + E_j^{(0)} e^{-i(k_g z -\omega_g t)})\,e^{i(k_g z -\omega_g t)}$, where $Q_{1,2,3}$ are the first three quantities listed in \ref{To_be_matched}, respectively. Similarly, for $j=1,2,3$, $-\sqrt{-g}\,G^{j0}$ becomes $(Q_{j+3} + B_j^{(0)} e^{-i(k_g z -\omega_g t)})\,e^{i(k_g z -\omega_g t)}$, where $Q_{4,5,6}$ are the last three quantities listed in \ref{To_be_matched}, respectively. Since ${\bm E^{(0)}}$ and ${\bm B^{(0)}}$ represent the background electric and magnetic fields, respectively, discontinuities in some of their components at the boundaries imply the presence of surface charges and current densities there~\cite{Griffiths:1999}. Hence, the remaining quantities to be matched are $Q_j$, with $j=1,2,\dots, 6$, as listed in \ref{To_be_matched}.}
\begin{equation}\label{To_be_matched}
    \widetilde{e}_x - \,_E\mathcal{S}_1^+,\,\,\,\,\, \widetilde{e}_y + \,_E\mathcal{S}_2^-,\,\,\,\,\, \widetilde{e}_z + \mathcal{S}_E,\,\,\,\,\, \widetilde{b}_{x},\,\,\,\,\, \widetilde{b}_{y},\,\,\,\,\, \widetilde{b}_{z}.
\end{equation}
After matching the above quantities at $z= -R_1$ and $z=R_2$, we obtain expressions for $C_{1,3}$, $\widetilde{C}_{1,3}$, and $f_{1,2,3,4}$. These expressions are provided in \ref{Appendix: Remaining expressions 1}. We can now write the solutions in all three regions explicitly, without any unknown functions or constants.

Hence, the solution in region III is
\begin{multline}
    \widetilde{e}_x = \widetilde{b}_y = \frac{ik_g (R_1+R_2)}{2}\, \left\{A_+( B_y^{(0)}- E_x^{(0)}) 
    - A_\times ( B_x^{(0)}
    \right.\\\left.
    +  E_y^{(0)} ) 
    + \frac{A_l}{2}(B_y^{(0)} + E_x^{(0)}) - A_x E_z^{(0)} - A_y B_z^{(0)}\right\},
\end{multline}
\begin{multline}
    \widetilde{e}_y = -\widetilde{b}_x = \frac{ik_g(R_1+R_2)}{2} \left\{A_+( E_y^{(0)}+ B_x^{(0)}) 
    - A_\times ( E_x^{(0)}
    \right.\\\left.
    -  B_y^{(0)} ) 
    + \frac{A_l}{2}(E_y^{(0)} - B_x^{(0)}) + A_x B_z^{(0)} - A_y E_z^{(0)}\right\},
\end{multline}
\begin{equation}
    \widetilde{e}_z = 0, \qquad \widetilde{b}_z = 0.
\end{equation}
Note that the $z$-components ($z$ is GW propagation direction) of the background electromagnetic fields always appear alongside additional polarization terms in the solution, as mentioned in the last paragraph of \ref{Section: Formalism}. Hence, if we set the additional polarization terms to zero, the solution depends only on the components of the background electric and magnetic fields perpendicular to the GW propagation direction. This is the outcome of the classic inverse GZ effect~\cite{Boccaletti:1970pxw, Boccaletti:1971}.

It is also interesting to note that the $z$-dependent part of the region II solution (\ref{Eq_31}$-$\ref{Eq_36}) is carried over to region III, with $z$ replaced by the width of region II, $R_1 + R_2$. The solutions in region III are proportional to the width of region II, which represents the distance travelled by the GW through the background electromagnetic field. Hence, the square of the width of region II will appear in the induced EMW flux. Thus, the wider the region of the background electromagnetic field, the greater the induced EMW flux. We will illustrate this with an explicit example in \ref{Section: GW passing through astrophysical magnetic fields: neutron star magnetosphere}. 

Similarly, the components of the induced electric and magnetic fields in region III are directly proportional to the magnitude of the background electromagnetic field. This dependence is also reflected in the induced EMW flux. Hence, in general, the stronger the background electromagnetic field, the greater the induced EMW flux. The same statement generally holds for the amplitude of the incident GW. 

At the outset, we would like to mention
the multiplicative term $\{-1+e^{2 i k_g (R_1+R_2)}\}$ in the solution for region I (see the expressions for $C_1$ and $\widetilde{C}_1$ in \ref{Appendix: Remaining expressions 1}). For a GW of specific wavelength $\lambda_g$ satisfying
\begin{equation}
    \frac{n \lambda_g}{2} = R_1 + R_2, \qquad n\in \mathbb{N}_0,
\end{equation}
the solution in region I vanishes, i.e., there will not be any EM wave, as reported by Ref.~\cite{Boccaletti:1970pxw}. Interestingly, this property holds even in the presence of the additional GW polarization modes.

\subsection{Time-independent magnetic field}\label{Section: Time-independent magnetic field}

In the previous section, we illustrated our formalism using the classic setting of the inverse GZ effect in the presence of a constant background electromagnetic field. In this section, we will demonstrate the applicability of our formalism to a time-independent, spatially varying magnetic field. Let us consider the components of the background electromagnetic field as follows:
\begin{equation}
    B_y^{(0)}(x,y,z) = \alpha z, \qquad B_z^{(0)}(x,y,z) = \alpha y,
\end{equation}
and
\begin{equation}
    E_{x,y,z}^{(0)}(x,y,z) = 0 = B_{x}^{(0)}(x,y,z).
\end{equation}
They satisfy the background Maxwell equations (\ref{BM1} and \ref{BM2}). Since the components of the background electromagnetic fields depend only on the spatial coordinates $y$ and $z$ and the incident GW propagates along the positive $z$ direction, we can replace the unknown functions of $x,y$ with functions of $y$ only, i.e., $f_{1,2,3,4}(x,y) = f_{1,2,3,4}(y)$. As in the previous section and in \ref{Footnote_2}, we can also express the unknown functions of $x,y,z$ as
\begin{equation}
    \phi_j(x,y,z) = \phi_{j0}(y, z) + x\,\phi_{j1}(z);\,\,\,\, j = 1,2.
\end{equation}

Now, we can calculate $\widetilde{\mathcal{Q}}_1(x,y,z)$ and $\widetilde{\mathcal{Q}}_2(x,y,z)$. The first constraint equations in \ref{Condition_1} and \ref{Condition_2} imply
\begin{equation}
    \phi_{21}(z) = 0, \qquad \phi_{11}(z) = 0,
\end{equation}
respectively.

The second equations in \ref{Condition_1} and \ref{Condition_2} provide second-order differential equations for $f_3(y)$ and $f_2(y)$, respectively, in terms of the other unknown functions. After analysing these equations, we finally obtain integro-differential equations for $\phi_{10}(y,z)$ and $\phi_{20}(y,z)$, which can be expressed collectively as a single equation:
\begin{equation}\label{IDE}
    \partial_y \phi_{j0}(y,z) + 2ik_g\int\mathrm{d}y\,\partial_z \phi_{j0}(y,z) + \int\mathrm{d}y\,\partial_z^2 \phi_{j0}(y,z) = 0,
\end{equation}
where $j=1,2$. The intermediate steps are detailed in \ref{Appendix: Remaining expressions 2a}.

To solve the integro-differential equations, let us consider the ansatz $\phi_{j0}(y,z) = U_j(y) + V_j(z)$, which implies
\begin{equation}
    -\frac{U_j'(y)}{y} = V_j''(z) + 2ik_gV_j'(z).
\end{equation}
Since the LHS is a function of $y$ and the RHS is a function of $z$, both must be equal to a constant (say, $C_{j0}$). This finally yields
\begin{equation}
    \phi_{j0}(y,z) = C_{j1}-\frac{C_{j0}\,y^2}{2}+ \frac{C_{j0}\,z}{2ik_g}-\frac{C_{j2}\,e^{-2ik_gz}}{2ik_g},
\end{equation}
where $C_{j1}$ and $C_{j2}$ are constants of integration. We can now write the forms of the other functions and the solution. The expressions for these unknown functions are provided in \ref{Appendix: Remaining expressions 2b}. This finally provides us with one of the solutions to this problem.

This section illustrated the formalism presented in \ref{Section: Formalism} through a spatially varying, time-independent magnetic field. The background field configurations considered in this and the preceding section might be useful for laboratory investigations and could also open up new prospects for GW detectors. Having illustrated the formalism with specific magnetic fields, we will now apply it to an astrophysical magnetic field in the next section.

\subsection{Astrophysical magnetic field}\label{Section: GW passing through astrophysical magnetic fields: neutron star magnetosphere}
In this section, we will utilise the formalism presented in \ref{Section: Formalism} to study an example in high-energy astrophysics. We will illustrate the inverse GZ effect when a GW passes through the magnetosphere of a star (mainly a neutron star (NS) or magnetar, although the effect can also be applied to other stellar magnetospheres). The electromagnetic field of a star's magnetosphere can convert a GW passing through it into an EMW (see also Ref.~\cite{Kushwaha:2022twx}). In this section, we will study these electromagnetic signals.

Let us consider a neutron star that is rotating about the $y$ axis, with its magnetic moment $\bm{\mu}$ aligned with its angular velocity $\bm{\Omega}$. The dipole model of the NS magnetosphere~\cite{Deutsch:1955, Rezzolla:2000dk, Cerutti:2016ttn} suggests that the magnetic field decays as $r^{-3}$ outside the NS, where $r$ is the radial distance from its centre.\footnote{Other models of the NS magnetosphere exist~\cite{Melrose:2016kaf}. Here, we use the dipole model to discuss the fall-off of the magnetic field of a neutron star. The dipole model may not be applicable near the surface of a neutron star~\cite{Domcke:2025qlw}.} For illustrative purposes, we assume that the NS magnetosphere is spherical and extends up to the light-cylinder radius $R_{\rm LC}$, which is related to the angular velocity by $R_{\rm LC} = c/\Omega$~\cite{Cerutti:2016ttn}.\footnote{It is important to mention that there is no hard boundary to the magnetosphere of a star. The electric and magnetic fields also exist outside this radius. However, the fields are very weak. In particular, the light-cylinder radius encompasses the inner magnetosphere of the NS (see Ref.~\cite{SKAPulsarScienceWorkingGroup:2025kss}).\label{Footnote_5}} We also assume a constant magnetic field in the NS magnetosphere, directed along the negative $y$-axis. Hence, the magnetic field configuration is
\begin{equation}
    B_y^{(0)}(x,y,z)= -B_y^{(0)},\,\,\,\, B_{x,z}^{(0)}(x,y,z)= 0 = E_{x,y,z}^{(0)}(x,y,z).
\end{equation}
The dipole model of the NS magnetosphere also proposes a non-zero electric field in the NS magnetosphere, which decays as $r^{-4}$~\cite{Cerutti:2016ttn}. For illustrative purposes, we consider only the magnetic field here. 

Let us consider a gravitational wave propagating along the positive $z$-axis and passing through the NS magnetosphere. Now, following the formalism presented in \ref{Section: Formalism} and the first example presented in \ref{Section: Constant background electric and magnetic field}, the induced electromagnetic field components resulting from the inverse GZ effect will be (see \ref{Separation}) 
\begin{equation}\label{Eq_Star_GR_1}
    \widetilde{E}_x = \widetilde{B}_y = -\frac{i k_g (2A_+ + A_l) B_y^{(0)} z_0}{4} e^{i(k_g z -\omega_g t)},
\end{equation}
\begin{equation}\label{Eq_Star_GR_2}
    \widetilde{E}_y = -\widetilde{B}_x = -\frac{i k_g A_\times B_y^{(0)} z_0}{2} e^{i(k_g z -\omega_g t)},
\end{equation}
and
\begin{equation}\label{Eq_Star_GR_3}
    \widetilde{E}_z = 0 = \widetilde{B}_z.
\end{equation}
where $z_0$ is the distance travelled by the GW in the stellar magnetosphere. Following the notation of \ref{Section: Constant background electric and magnetic field}, these are region III solutions. It is interesting to note that, although we consider all extra polarizations of the GW, only the longitudinal mode affects the induced EMW when only the $y$ component of the background magnetic field is present. In addition, in the absence of extra polarization terms, the plus mode induces the $x~(y)$ component of the electric (magnetic) field, while the cross mode induces the $y~(x)$ component of the electric (magnetic) field.

\begin{figure}[h!]
    \centering
    \hspace{-0.3in}\includegraphics[scale=0.6]{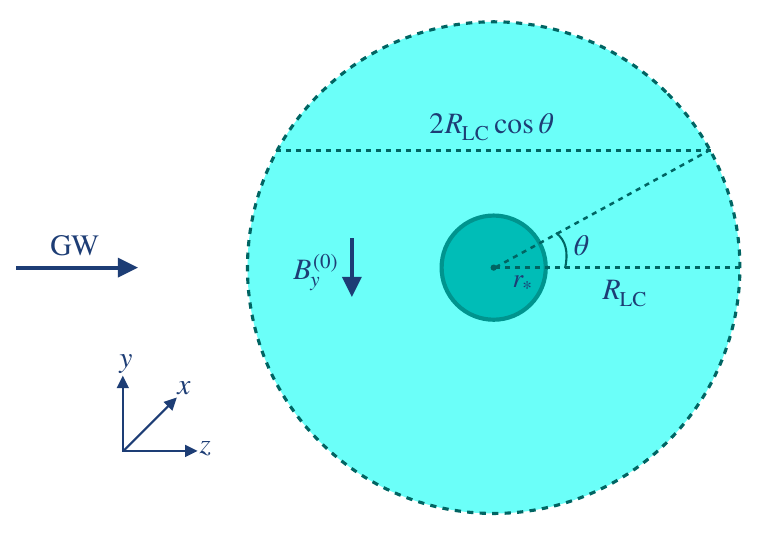}
    \caption{Approximate schematic representation of a star's magnetosphere. The radius of the star is denoted by $r_*$. The radius of the star's magnetosphere is given by the light-cylinder radius $R_{\rm LC}$. The GW propagates along the positive $z$-direction. Hence, it travels different distances through different parts of the star's magnetosphere, with the distance given by $2R_{\rm LC}\cos\theta$. The magnetic field points in the negative $y$-direction. The coordinate system is also shown in the bottom-left corner, where the $y$- and $z$-axes lie in the plane of the paper, and the $x$-axis points inward, perpendicular to the plane of the paper.}
    \label{Figure:Schematic description of a star's magnetosphere}
\end{figure}

In this example of a stellar magnetic field, we consider the magnetosphere to be spherical. In addition, we assume that the radius of the star is much smaller than the light-cylinder radius, $r_*\ll R_{\rm LC}$, or equivalently, $\Omega\,r_*/c\ll 1$. Thus, along the direction of propagation, the GW travels a distance of $2 R_{\rm LC} \cos \theta$ through the magnetosphere, where $\theta$ is the angle between the line connecting the centre of the star to a point on the {\em surface} of the magnetosphere and the $z$-axis. The origin of the coordinate system is at the centre of the star. Therefore, the components of the induced electric and magnetic fields emanating from that point on the surface are the same as those given in \ref{Eq_Star_GR_1}$-$\ref{Eq_Star_GR_3}, with $z_0$ replaced by $2 R_{\rm LC} \cos\theta$ (see \ref{Figure:Schematic description of a star's magnetosphere}).

Since the GW is converted into an EMW, we can quantify this conversion by taking the ratio of the energy density of the induced EMW ($\rho_{\scriptscriptstyle\rm EMW}$) to that of the incident GW ($\rho_{\scriptscriptstyle\rm GW}$). We call this quantity the conversion factor, $\alpha = \rho_{\scriptscriptstyle\rm EMW}/\rho_{\scriptscriptstyle\rm GW}$. 

Since extra modes appear in modified theories of gravity and the theories can differ from one another by construction, a universal expression for the energy density of GWs cannot be established without specific knowledge of the theory. Thus, for computational purposes, we write the energy density of GWs in GR in terms of only the $+$ and $\times$ polarization modes, which is given by~\cite{Misner:1973prb}
\begin{equation}
    \rho_{\scriptscriptstyle\rm GW}^{+\times} = \frac{c^2 \omega_g^2}{32\pi G}\,\left(|A_+|^2 + |A_\times|^2\right),
\end{equation}
where we have used 
\begin{equation}
    A_j = |A_j|\,e^{i\delta_j};\qquad j=+,\times,b,l,x,y.
\end{equation}
$\delta_j$ represents the contribution to the phase term of the corresponding GW mode and $|A_j|$ is the amplitude of that mode. However, we can anticipate that the effects of the additional polarization modes will be similar to those of the $+$ and $\times$ modes. The effects of phase differences between different GW polarization modes may also appear in this expression.

The energy density of the EMW is~\cite{Jackson:1998nia}
\begin{equation}
    \rho_{\scriptscriptstyle\rm EMW} = \frac{|\bm{\widetilde{E}}\cdot\bm{\widetilde{E}^*}|+|\bm{\widetilde{B}}\cdot\bm{\widetilde{B}^*}|}{16\pi},
\end{equation}
where the superscript $*$ denotes the complex conjugate. In this example, the energy density of the induced EMW becomes
\begin{equation}
    \rho_{\scriptscriptstyle\rm EMW}(\theta) = \frac{k_g^2 (B_y^{(0)})^2 R_{LC}^2\cos^2\theta}{8\pi} \,\mathcal{C}_{\scriptscriptstyle\rm EMW},
\end{equation}
where
\begin{equation}
    \mathcal{C}_{\scriptscriptstyle\rm EMW} = |A_+|^2  + |A_\times|^2 
    + \frac{|A_l|^2}{4} +  |A_+||A_l|\cos{(\delta_+-\delta_l)},
\end{equation}
where $\delta_+-\delta_l$ is the phase difference between the plus ($+$) and longitudinal ($l$) polarization modes.

Since we have considered only the $+$ and $\times$ modes in calculating the energy density of the GW, we will consider only these modes when computing the conversion factor. Thus, the conversion factor will be 
\begin{equation}
    \alpha^{+,\times}(\theta) = \frac{4G(B_y^{(0)})^2 R_{\rm LC}^2\cos^2\theta}{c^4}.
\end{equation}
This is the conversion factor for the EMW emanating from a particular point of the star's magnetosphere (where the incident GW has only $+$ and $\times$ modes). It is interesting to note that, when only the $+$ and $\times$ modes are considered, the conversion factor does not depend on the GW parameters. However, with extra polarizations, we expect the GW energy density to depend explicitly on the GW amplitudes and the phase differences between different modes. In that case, the conversion factor might depend on the GW parameters. We can calculate the total conversion factor by integrating over the magnetosphere
\begin{equation}
    \alpha_{\rm tot}^{+\times} = \frac{4G(B_y^{(0)})^2 R_{\rm LC}^2}{3c^4}.
\end{equation}
The total conversion factor is directly proportional to the square of the magnetic field and the square of the radius of the star's magnetosphere. This means that the stronger the magnetic field and the larger the radius of the magnetosphere, the greater the conversion.

Now, let us compute the Poynting vector of the induced EMW. The time-averaged (averaged over one cycle) Poynting vector is given by~\cite{Jackson:1998nia}
\begin{equation}
    \langle{\bm{S}}\rangle = \frac{c}{8\pi}{\rm Re}(\bm{\widetilde{E}}\times \bm{\widetilde{B}}^*),
\end{equation}
where $\langle\rangle$ denotes the time average. Since the $x$- and $y$-components of the induced electric and magnetic fields are the only non-zero components, the Poynting vector points along the $z$-axis, which is also the direction of the incident GW. Hence, the time-averaged Poynting vector of the EMW emanating from a particular point (identified by $\theta$, see \ref{Figure:Schematic description of a star's magnetosphere}) from the NS magnetosphere is given by
\begin{equation}
    \langle{\bm{S}}\rangle = \frac{(B_y^{(0)})^2 \omega_g^2 R_{\rm LC}^2 \cos^2\theta}{8\pi c}\left(\frac{|2A_+ 
    +A_l|^2}{4} + |A_\times|^2\right)\,\hat{\bm{z}}.
\end{equation}
Now, as we did for the conversion factor, the total (angle-averaged) Poynting flux will be
\begin{equation}
    \langle {\bm{S}} \rangle_{\rm tot} = \frac{(B_y^{(0)})^2\omega_g^2 R_{\rm LC}^2}{24\pi c}\,\mathcal{C}_{\scriptscriptstyle\rm EMW}\,\hat{\bm{z}}.
\end{equation}

Apart from the magnetic field and the radius of the magnetosphere, the Poynting flux also depends on the GW parameters, namely, amplitude, frequency, and phase difference between $+$ and $l$ modes. However, the phase-difference term does not appear when we consider only the $+$ and $\times$ GW polarization modes.

Since the induced EMW is monochromatic, the equivalent ``spectral flux density (SFD)'' of this signal can be defined as the ratio of the magnitude of time-averaged Poynting flux to the EMW frequency ($f_g = \omega_g/2\pi$), given by~\footnote{The jansky (Jy) is the unit of SFD: 1 Jy = $10^{-23}\,$erg s$^{-1}$ cm$^{-2}$ Hz$^{-1}$.}
\begin{equation}
    \hspace{-0.25in}
    \frac{\langle S \rangle_{\rm tot}}{f_g} = \mathcal{X}\left(\frac{|A|}{10^{-23}}\right)^2 \left(\frac{B_y^{(0)}}{10^{10}~{\rm G}}\right)^2 \left(\frac{f_g}{1~{\rm MHz}}\right) \left(\frac{R_{\rm LC}}{10^8~{\rm cm}}\right)^2 ~{\rm Jy},
\end{equation}
where the multiplicative factor $\mathcal{X}$ takes values
\begin{equation}
    \mathcal{X} = 
    \begin{cases}
		3.49\times 10^8, & |A_{+,\times}| = |A|, \,\,\, |A_l| = 0,\\
        3.93\times 10^8, & |A_{+,\times,l}| = |A|,\,\,\,\delta_+-\delta_l = \pi/2,\\
        5.68\times 10^8, & |A_{+,\times,l}| = |A|,\,\,\,\delta_+-\delta_l = 0.
	\end{cases}
\end{equation} 

We assumed a constant magnetic field in the NS magnetosphere. However, proposed models of the neutron star magnetosphere (e.g., the dipole model) suggest that the NS magnetic field decreases as $r^{-3}$ with distance from its surface. Hence, we can approximate it by a constant value by averaging it over the entire magnetosphere, which yields
\begin{equation}
    B_y^{(0)} \approx  \frac{3 B_* r_* ^3}{R_{\rm LC}^3}\ln\left(\frac{R_{\rm LC}}{r_*}\right),
\end{equation}
where we have considered the magnetic field in the NS magnetosphere as $B_* (r_*/r)^3$, with $B_*$ being the magnetic field at the surface of the NS. As stated before, we are also assuming $R_{\rm LC} \gg r_*$. Now, we can write the SFD of the induced EMW as
\begin{multline}
    \hspace{-0.1in}
    \frac{\langle S \rangle_{\rm tot}}{f_g} \approx  \mathcal{Y}\left(\frac{|A|}{10^{-23}}\right)^2 \left(\frac{B_*}{10^{12}~{\rm G}}\right)^2 \left(\frac{f_*}{10~{\rm Hz}}\right)^4 
    \left(\frac{f_g}{1~{\rm MHz}}\right)\\
    \times \left(\frac{r_*}{10~{\rm km}}\right)^6 \left[\ln{\left\{477.13\times\left(\frac{10~{\rm Hz}}{f_*}\right)\left(\frac{10~{\rm km}}{r_*}\right)\right\}}\right]^2~{\rm Jy},
\end{multline}
where $f_* = \Omega/2\pi$ and the multiplicative factor $\mathcal{Y}$ is
\begin{equation}
    \mathcal{Y} = 
    \begin{cases}
		6.07\times 10^{-2}, & |A_{+,\times}| = |A|, \,\,\, |A_l| = 0,\\
        6.82\times 10^{-2}, & |A_{+,\times,l}| = |A|,\,\,\,\delta_+-\delta_l = \pi/2,\\
        9.86\times 10^{-2}, & |A_{+,\times,l}| = |A|,\,\,\,\delta_+-\delta_l = 0.
	\end{cases}
\end{equation}
Now, if a GW without any extra polarizations, with $|A_+| = 10^{-23} = |A_\times|$ and $f_g = 1$ MHz, passes through a typical NS with a surface magnetic field of $10^{12}$ G, a radius of $10$ km, and a rotational frequency of $10$ Hz, the SFD is $2.31$ Jy~\cite{Heiselberg:478284}. For the same configuration, if we take extra GW polarization modes into account ($|A_{+,\times,l}| = 10^{-23}$) and consider the phase difference $\delta_+-\delta_l = 0$, the SFD is $3.75$ Jy. Hence, the effect is very small. Since the SFD is linearly proportional to the GW frequency ($\omega_g$), for a GW with a frequency of $1$ GHz, the SFD would be $1000$ times the previous values, making the absolute difference significant. However, the same SFD can be achieved with a different magnetic field. Note that the induced EMW has the same frequency as the incident GW.

\begin{figure}[h!]
    \centering
    \includegraphics[scale=0.5]{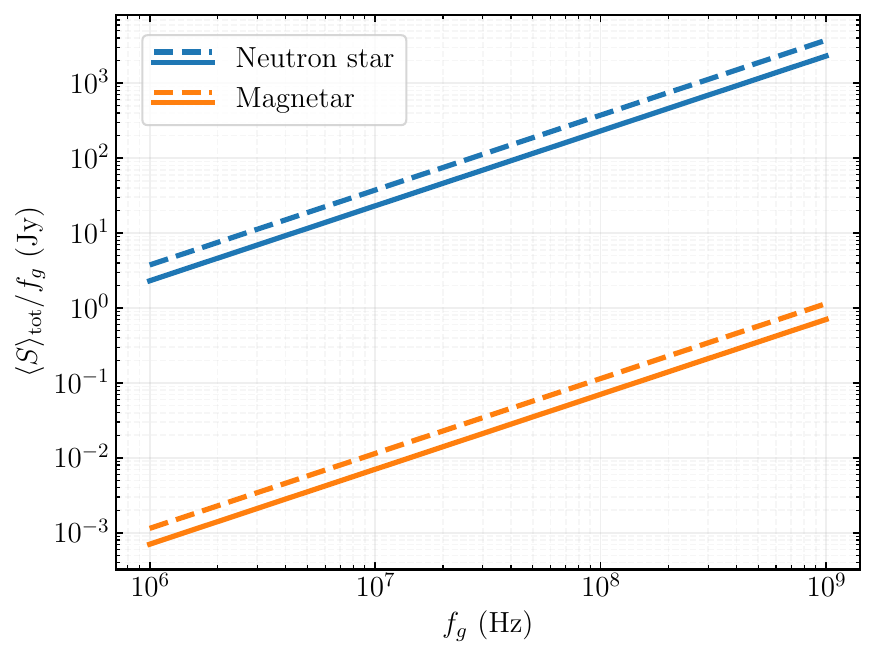}
    \caption{The spectral flux density (SFD) of the induced EMW is plotted against the frequency of the incident GW. The blue solid and dashed lines correspond to a typical neutron star with a radius $(r_*)$ of $10$ km, a surface magnetic field $(B_*)$ of $10^{12}$ G, and a rotational frequency $(f_*)$ of $10$ Hz, while the orange solid and dashed lines correspond to a typical magnetar with a radius $(r_*)$ of $10$ km, a surface magnetic field $(B_*)$ of $10^{14}$ G, and a rotational frequency $(f_*)$ of $0.1$ Hz~\cite{Heiselberg:478284}. For both cases, the solid lines correspond to GWs without extra polarization modes (i.e., with only the plus and cross polarization modes), while the dashed lines correspond to GWs with extra polarization modes and zero phase difference between the plus and longitudinal modes. In all cases, the amplitude of the incident GW is assumed to be $10^{-23}$. The SFD ranges are similar to the observed peak Poynting fluxes of fast radio bursts~\cite{CHIMEFRB:2025ggb}. Note that this analysis is approximate and assumes a constant magnetic field throughout the NS magnetosphere.} 
    \label{Figure:Spectral flux density vs GW frequency}
\end{figure}

The SFD has also been plotted against the incident GW frequency in \ref{Figure:Spectral flux density vs GW frequency}, for a typical NS and a typical magnetar. It is interesting to note that the range of SFDs is around $0.001-1000$ Jy. The SFD can also be higher or lower than this range. This range is similar to the observed peak Poynting fluxes of fast radio bursts~\cite{CHIMEFRB:2025ggb}. 

It would be interesting to explore the outcome of the inverse GZ effect using rigorous models of the NS magnetosphere. The formalism presented in this paper could be important for such a study because models such as the dipole model of the NS magnetosphere propose time-independent, spatially varying electromagnetic fields in the NS magnetosphere. It is worthwhile to point out the difference in the SFDs of the induced EMW from a typical NS (or magnetar) due to the extra GW polarization modes, although similar values can also be achieved without extra polarization modes. Thus, with the current model of a constant magnetic field in the NS magnetosphere, this cannot be considered a definitive test for the presence or absence of extra polarization modes. However, with rigorous models of the NS magnetosphere, the difference in the SFDs due to the presence and absence of extra polarization modes could offer a way to test for these modes.

Having calculated the electromagnetic signal emanating from the stellar magnetosphere due to the inverse GZ effect, it would be interesting to study its polarization properties. To analyse these properties, we calculate the Stokes parameters~\cite{Stokes:1851,Rybicki_Lightman:1979,Qu:2023ibp,Qu:2026pjv}:
\begin{equation}
    I = \frac{\widetilde{E}_x\,\widetilde{E}_x^* + \widetilde{E}_y\,\widetilde{E}_y^*}{2}, \qquad Q = \frac{\widetilde{E}_x\,\widetilde{E}_x^* - \widetilde{E}_y\,\widetilde{E}_y^*}{2},
\end{equation}
\begin{equation}
    U = {\rm Re}(\widetilde{E}_x\,\widetilde{E}_y^*), \qquad V = -{\rm Im}(\widetilde{E}_x\,\widetilde{E}_y^*),
\end{equation}
where $I$ is the time-averaged total intensity and $\sqrt{Q^2 + U^2 + V^2}$ is the time-averaged intensity of the polarized light ($I_{\rm pol}$). To further specify the polarization, we define the following three ratios,
\begin{equation}
    \Pi = \frac{I_{\rm pol}}{I}, \qquad \Pi_{\rm L} = \frac{\sqrt{Q^2 + U^2}}{I}, \qquad \Pi_{\rm V} = \frac{|V|}{I},
\end{equation}
where the first ratio, $\Pi$, defines the degree of polarization. If $\Pi = 1$, the light is completely polarized. The second ($\Pi_{\rm L}$) and third ($\Pi_{\rm V}$) ratios define the linear and circular degrees of polarization, respectively. $\Pi_{\rm L} = 1(0)$ or $\Pi_{\rm V} = 0(1)$ indicate that the light is completely linearly (circularly) polarized~\cite{Stokes:1851,Rybicki_Lightman:1979,Qu:2023ibp,Qu:2026pjv}.  

Now, we can calculate all the quantities for the EM signal emanating from a particular point (identified by $\theta$; see \ref{Figure:Schematic description of a star's magnetosphere}) from the NS magnetosphere and obtain
\begin{equation}
    I =  \sqrt{Q^2 + U^2 +V^2}, \qquad {\rm or} \qquad \Pi = 1,
\end{equation}
which means that the light is completely polarized. Similarly, we can calculate other quantities. The expressions for the Stokes parameters are provided in \ref{Appendix: Remaining expressions 3}. 

We can also calculate $\Pi_{\rm L}$ and $\Pi_{\rm V}$ from the Stokes parameters. Since identifying the overall polarization of GWs in the presence of extra modes is not well understood~\cite{Isi:2022mbx, Anile:1974}, we present $\Pi_{\rm L}$ and $\Pi_{\rm V}$ for the induced EM waves when the incident GW comprises only the $+$ and $\times$ polarization modes. In this case, $\Pi_{\rm L}$ and $\Pi_{\rm V}$ are
\begin{equation}
    \Pi_{\rm L}= \sqrt{1 -\frac{4|A_+|^2|A_\times|^2}{\left(|A_+|^2 + |A_\times|^2\right)^2}\sin^2(\delta_+-\delta_\times)},
\end{equation}
and
\begin{equation}
    \Pi_{\rm V} = \frac{2|A_+||A_\times|}{|A_+|^2 + |A_\times|^2}|\sin(\delta_+-\delta_\times)|.
\end{equation}
Note that if $|A_+|$ or $|A_\times|$ is zero, the incident GW is linearly polarized. In this case, $\Pi_{\rm L}= 1$ and $\Pi_{\rm V} = 0$, which means that the induced EMW will also be linearly polarized. Similarly, if $|A_+| = |A_\times|$ and $\delta_+ - \delta_\times = (2n+1)\pi/2$, where $n\in\mathbb{N}_0$, the incident GW is circularly polarized. In this case, $\Pi_{\rm L}= 0$ and $\Pi_{\rm V} = 1$, implying that the induced EMW will also be circularly polarized. Hence, linearly (circularly) polarized GWs with $+$ and $\times$ polarization modes induce linearly (circularly) polarized EMWs.

\section{Discussion}\label{Section: Discussion}

The inverse Gertsenshteǐn-Zel'dovich (GZ) effect connects two essential pillars of multi-messenger astronomy--gravitational waves (GWs) and electromagnetic waves (EMWs)--through a background electromagnetic field. It has numerous applications in astrophysics, ranging from the development of high-frequency gravitational wave (HFGW) detectors to the generation of astrophysical electromagnetic signals arising from the passage of GWs through astrophysical electromagnetic fields. In this work, we have presented a general formalism for calculating the components of the induced electric and magnetic fields resulting from the passage of a gravitational wave, including all possible extra polarization modes, through a spatially varying, time-independent background electromagnetic field. We have derived expressions for the components of the induced electric and magnetic fields and illustrated the formalism through several examples and applications. Thus, the formalism developed in this paper could provide an important framework for a wide range of applications involving the inverse GZ effect.

First, the formalism presented in this work can be employed in laboratory-type setups to investigate the effects of GWs on electromagnetic fields, potentially opening up new possibilities for GW detection. Second, astrophysical electromagnetic signals associated with the inverse GZ effect may provide a possible explanation for certain observed or yet-unidentified electromagnetic transients. In \ref{Section: GW passing through astrophysical magnetic fields: neutron star magnetosphere}, we analysed the electromagnetic signals associated with the inverse GZ effect in neutron star (NS) and magnetar magnetospheres, assuming a constant background magnetic field. The results are particularly interesting, as the induced electromagnetic waves exhibit a spectral flux density (SFD) comparable to that of fast radio bursts (FRBs). This motivates extending the analysis to more realistic models of NS magnetospheres. Since established models of NS magnetospheres, e.g., the dipole model~\cite{Rezzolla:2000dk, Cerutti:2016ttn}, also propose time-independent, spatially varying electromagnetic fields, our formalism could be particularly useful for such investigations. However, the electromagnetic field structure in realistic NS magnetospheres is considerably more complex. Therefore, numerical methods would likely be necessary to investigate the resulting electromagnetic signals in detail.

In addition, we have explored the effects of extra GW polarization modes on the electromagnetic waves induced through the inverse GZ effect. We have shown that the presence of extra GW polarizations can affect the SFD of the induced electromagnetic waves. However, within our present model of a constant magnetic field in the NS magnetosphere, it is not possible to establish with certainty whether a particular SFD provides evidence for the presence or absence of extra GW polarization modes, since similar SFDs can arise from different magnetic field configurations in the NS magnetosphere. Nevertheless, by employing more realistic models of NS magnetospheres, we expect that our formalism could provide more accurate predictions for the induced electromagnetic signals and potentially help distinguish the effects of extra GW polarizations from those arising from the background magnetic field. Such an analysis could therefore offer a possible avenue for testing the presence or absence of extra polarization modes in gravitational waves.

At this stage, it is important to discuss the applicability of our formalism to incident GWs with extra polarization modes propagating at different speeds. As discussed in \ref{Section: Effects of the speed of GW propagation}, our formalism can also be employed to construct solutions when different polarization modes of an incident GW propagate with different speeds. In the applications considered in this work, however, we have assumed that all polarization modes propagate at the speed of light in vacuum. The consequences of different propagation speeds for different GW polarization modes require a more detailed investigation, which we leave for future work. Nevertheless, it is useful to discuss a possible observational implication in the context of the inverse GZ effect. As studied in Ref.~\cite{Schumacher:2023jxq}, even a minute difference in the propagation speeds of different GW polarization modes can produce measurable differences in their arrival times at a detector, owing to the large distances between astrophysical GW sources and observers, typically of megaparsec scales or larger. We expect that a related effect could arise in the inverse GZ process. If different GW polarization modes propagate at different speeds, the faster modes may induce electromagnetic waves earlier than the slower modes, potentially producing a temporal separation between the corresponding electromagnetic signals. Such a temporal separation could provide a possible observational signature of extra GW polarization modes. However, this possibility requires a separate and detailed investigation, particularly to determine whether the resulting time delays could be distinguished from those arising from other astrophysical mechanisms or propagation effects.

Finally, the background electromagnetic field considered in this work has been assumed to be time-independent. In general, however, astrophysical electromagnetic fields can depend on both spatial coordinates and time. It would therefore be interesting, although technically more challenging and non-trivial, to extend our formalism to a background electromagnetic field that varies in both space and time and derive the corresponding components of the induced electric and magnetic fields. Since electromagnetic fields in stellar magnetospheres are generally time-dependent, such an extension would allow us to investigate the inverse GZ effect in more realistic astrophysical environments. Another important direction for future research would be to incorporate plasma effects, including dissipative properties such as viscosity, into the analysis. This would also require examining the role of the coherence length in the conversion process~\cite{Zeldovich:1974}, which could significantly influence the strength and propagation of the induced electromagnetic signals. A further extension would be to investigate the Maxwell equations in more general curved spacetimes, rather than restricting the analysis to perturbations around Minkowski spacetime. Such developments would broaden the applicability of the present formalism and provide a more comprehensive understanding of the inverse GZ effect in realistic astrophysical settings.

\begin{acknowledgments}
RV thanks IIT Gandhinagar Director's fellowship to support his research. RP acknowledges support from the Achievers Postdoctoral Fellowship at IIT Gandhinagar. S.S.’s research is supported by the Department of Science and Technology, Government of India, under the ANRF CRG Grant (No. CRG/2023/000934).

\end{acknowledgments}
\onecolumngrid
\vspace{0.25in}
\rule{\textwidth}{1pt}
\appendix
\labelformat{section}{Appendix #1} 
\labelformat{subsection}{Appendix \thesection.#1}
\labelformat{subsubsection}{Appendix \thesection.\thesubsection.#1}
\vspace{0.25in}
\section{Remaining expressions from \ref{Section: Formalism}}\label{Appendix: Some other expressions}
In this appendix, we provide a few remaining expressions from \ref{Section: Formalism} of the main text. First, we provide the expressions for all the components of $\bm{\widetilde{e}}$ and $\bm{\widetilde{b}}$. Then, we present the mathematical forms of the quantities $\widetilde{\mathcal{Q}}_1$ and $\widetilde{\mathcal{Q}}_2$, which are essential for obtaining the constraint equations for the unknown functions appearing in $\bm{\widetilde{e}}$ and $\bm {\widetilde{b}}$.

The components of $\bm{\widetilde{e}}$ are
    \begin{multline}
        \widetilde{e}_x = \frac{f_3(x,y)\,e^{-ik^-z}}{2} + \frac{f_4(x,y)\,e^{-ik^+z}}{2} -\frac{i\omega_g}{2c}e^{-ik^-z}\int \mathrm{d}z\,e^{ik^-z} \left\{- \partial_x \phi_1 -\partial_y \phi_2 + \mathcal{G}_2\right\}
        \\
        -\frac{i\omega_g}{2c}e^{-ik^+z}\int \mathrm{d}z\,e^{ik^+z} \left\{- \partial_x \phi_1 +\partial_y \phi_2 - \mathcal{G}_2\right\},
    \end{multline}
    \begin{multline}
        \widetilde{e}_y = \frac{f_1(x,y)\,e^{-ik^+z}}{2} -\frac{f_2(x,y)\,e^{-ik^-z}}{2} -\frac{i\omega_g}{2c}e^{-ik^+z}\int \mathrm{d}z\,e^{ik^+z} \left\{-\partial_y \phi_1 - \partial_x \phi_2 +\mathcal{G}_1\right\}
    \\
    +\frac{i\omega_g}{2c}e^{-ik^-z}\int \mathrm{d}z\,e^{ik^-z} \left\{\partial_y \phi_1 - \partial_x \phi_2 + \mathcal{G}_1\right\},
    \end{multline}
    and
    \begin{multline}
    \widetilde{e}_z =
    \frac{ic\{\partial_x f_3(x,y)-\partial_y f_2(x,y)\}\,e^{-ik^-z}}{2\omega_g} - \frac{ic\{\partial_x f_4(x,y) + \partial_y f_1(x,y)\}\,e^{-ik^+z}}{2\omega_g} -\frac{ic(\partial_x \,_{B}\mathcal{S}^{-}_{2}
    +\partial_y \,_{B}\mathcal{S}^{+}_{1})}{\omega_g}
    -\mathcal{S}_E
    \\
    -\frac{1}{2}e^{-ik^-z}\int \mathrm{d}z\,e^{ik^-z} \left( \partial_x^2 \phi_1 +\partial_y^2 \phi_1 - \partial_x \mathcal{G}_2 + \partial_y\mathcal{G}_1\right)
    +\frac{1}{2}e^{-ik^+z}\int \mathrm{d}z\,e^{ik^+z} \left( \partial_x^2 \phi_1 +\partial_y^2 \phi_1 + \partial_x\mathcal{G}_2 - \partial_y\mathcal{G}_1\right).
\end{multline}
The components of $\bm{\widetilde{b}}$ are
    \begin{multline}
        \widetilde{b}_x = \frac{f_1(x,y)\,e^{-ik^+z}}{2} +\frac{f_2(x,y)\,e^{-ik^-z}}{2} -\frac{i\omega_g}{2c}e^{-ik^+z}\int \mathrm{d}z\,e^{ik^+z} \left\{-\partial_y \phi_1 - \partial_x \phi_2 + \mathcal{G}_1\right\}
    \\
    -\frac{i\omega_g}{2c}e^{-ik^-z}\int \mathrm{d}z\,e^{ik^-z} \left\{\partial_y \phi_1 - \partial_x \phi_2 + \mathcal{G}_1\right\},
    \end{multline}
    \begin{multline}
        \widetilde{b}_y = \frac{f_3(x,y)\,e^{-ik^-z}}{2} - \frac{f_4(x,y)\,e^{-ik^+z}}{2} -\frac{i\omega_g}{2c}e^{-ik^-z}\int \mathrm{d}z\,e^{ik^-z} \left\{- \partial_x \phi_1 -\partial_y \phi_2 + \mathcal{G}_2\right\}
        \\
        +\frac{i\omega_g}{2c}e^{-ik^+z}\int \mathrm{d}z\,e^{ik^+z} \left\{- \partial_x \phi_1 +\partial_y \phi_2 - \mathcal{G}_2\right\},
    \end{multline}
    and
    \begin{multline}
    \widetilde{b}_z = \frac{ic\{\partial_y f_3(x,y)+\partial_x f_2(x,y)\}\,e^{-ik^-z}}{2\omega_g} + \frac{ic\{\partial_y f_4(x,y)-\partial_x f_1(x,y)\}\,e^{-ik^+z}}{2\omega_g} 
    \\
    -\frac{1}{2}e^{-ik^-z}\int \mathrm{d}z\,e^{ik^-z} \left( \partial_x^2 \phi_2 +\partial_y^2 \phi_2 - \partial_x \mathcal{G}_1 - \partial_y \mathcal{G}_2\right)
    +\frac{1}{2}e^{-ik^+z}\int \mathrm{d}z\,e^{ik^+z} \left( \partial_x^2 \phi_2 +\partial_y^2 \phi_2 -\partial_x \mathcal{G}_1 - \partial_y \mathcal{G}_2\right).
\end{multline}
In these expressions 
\begin{equation}
    k^\pm = k_g \pm \frac{\omega_g}{c}, \qquad \mathcal{G}_1 = - \mathcal{D}\,_{B}\mathcal{S}^{+}_{1}
    +\,_{E}\mathcal{S}^{-}_{2}, \qquad {\rm and} \qquad \mathcal{G}_2 = \mathcal{D}\,_{B}\mathcal{S}^{-}_{2} + \,_{E}\mathcal{S}^{+}_{1}.
\end{equation}
The expressions of $\widetilde{\mathcal{Q}}_1(x,y,z)$ and $\widetilde{\mathcal{Q}}_2(x,y,z)$ are
\begin{multline}
    \widetilde{\mathcal{Q}}_1(x,y,z) = \frac{1}{2k_g\omega_g}\left(e^{-ik^+z}\int \mathrm{d}z\,e^{ik^+z}(\mathcal{V}_1+\mathcal{L}_2) - e^{-ik^-z}\int \mathrm{d}z\,e^{ik^-z}(\mathcal{V}_1+\mathcal{L}_2)\right)
    \\
    +\frac{ic}{2\omega_g}\left[e^{-ik^+z}\{\partial_y f_4(x,y)-\partial_xf_1(x,y)\}+ e^{-ik^-z}\{\partial_y f_3(x,y)+\partial_xf_2(x,y)\}\right]
    -\frac{i\omega_g \phi_2}{c}-S_B,
\end{multline}
and
\begin{multline}
    \widetilde{\mathcal{Q}}_2(x,y,z) = \frac{1}{2k_g\omega_g}\left(e^{-ik^+z}\int \mathrm{d}z\,e^{ik^+z}(\mathcal{V}_2-\mathcal{L}_1) + e^{-ik^-z}\int \mathrm{d}z\,e^{ik^-z}(\mathcal{V}_2+\mathcal{L}_1)\right)
    \\
    +\frac{ic}{2\omega_g}\left[e^{-ik^+z}\{\partial_y f_1(x,y)+\partial_xf_4(x,y)\}+ e^{-ik^-z}\{\partial_y f_2(x,y)-\partial_xf_3(x,y)\}\right]
    +\frac{i\omega_g \phi_1}{c}+S_E + \frac{c}{k_g\omega_g}\mathcal{V}_3,
\end{multline}
where
\begin{equation}
    \mathcal{V}_1 = -2ick_g(\partial_x R_+^{(1)} + \partial_y R_+^{(2)} )- c\partial_z [\partial_x(R_+^{(1)}-R_-^{(1)})
    + \partial_y (R_+^{(2)}-R_-^{(2)})],
\end{equation}
\begin{equation}
    \mathcal{V}_2 = -2ick_g(\partial_x R_+^{(2)} - \partial_y R_+^{(1)} )- c\partial_z [\partial_x(R_+^{(2)}-R_-^{(2)})
    - \partial_y (R_+^{(1)}-R_-^{(1)})],
\end{equation}
and
\begin{equation}
    \mathcal{V}_3 = \partial_x (R_+^{(2)} - R_-^{(2)} ) - \partial_y (R_+^{(1)} - R_-^{(1)}), \qquad \mathcal{L}_j = k_g\omega_g(\partial_x^2\phi_j + \partial_y^2\phi_j);\,\,\,\,\, j = 1,2.
\end{equation}
In these expressions, $\mathcal{R}_\pm^{(1,2)}$, $\,_{E,B}\mathcal{S}^{\pm}_{i}$, and $\mathcal{S}_{E,B}$ are functions of $x,y,$ and $z$, defined in the main text (\ref{Eq_11}$-$\ref{Eq_14}). Some of these quantities are related to each other by
\begin{equation}\label{Rel_1}
    \mathcal{R}_+^{(1)} + \mathcal{R}_-^{(1)} = -\frac{i\omega_g}{c} \,_E \mathcal{S}_2^-, \qquad \mathcal{R}_+^{(1)} - \mathcal{R}_-^{(1)} = -i k_g \,_B \mathcal{S}_1^+, \qquad \mathcal{R}_+^{(2)} + \mathcal{R}_-^{(2)} = -\frac{i\omega_g}{c} \,_E \mathcal{S}_1^+, \qquad \mathcal{R}_+^{(2)} - \mathcal{R}_-^{(2)} = i k_g \,_B \mathcal{S}_2^-.
\end{equation}

\section{Remaining expressions from \ref{Section: Applications}}\label{Appendix: Remaining expressions from application section}
In this appendix, we provide a few remaining equations and expressions from \ref{Section: Applications} that we have not explicitly written in the main text.
\subsection{Remaining expressions from \ref{Section: Constant background electric and magnetic field}}\label{Appendix: Remaining expressions 1}
In this section, we will write down a few remaining expressions from \ref{Section: Constant background electric and magnetic field}. The expressions for $C_1$ and $\widetilde{C}_1$, which constitute the region I solutions, are
\begin{equation}
    C_1 = \frac{i\,\mathcal{R}_-^{(2)} \left(-1+e^{2 i k_g (R_1+R_2)}\right)e^{-2 i k_g R_1}}{2 k_g}, \qquad \widetilde{C}_1 = -\frac{i\, \mathcal{R}_-^{(1)}\,e^{-2 i k_g R_1} \left(-1+e^{2 i k_g (R_1+R_2)}\right) }{2 k_g}.
\end{equation}
The expressions for $C_3$ and $\widetilde{C}_3$, which fix the region III solutions, are
\begin{equation}
    C_3 = (R_1+R_2)\,\mathcal{R}_+^{(2)}, \qquad \widetilde{C}_3 = -(R_1+R_2)\,\mathcal{R}_+^{(1)}.
\end{equation}
The expressions for $f_1$, $f_2$, $f_3$, and $f_4$ from the region II solutions are
\begin{equation}
    f_1 = -\frac{i\,\mathcal{R}_-^{(1)}\,e^{2 i k_g R_2}}{k_g}, \qquad f_2 = \,_E\mathcal{S}_2^- + 2R_1\mathcal{R}_+^{(1)}, \qquad f_3 = \,_E\mathcal{S}_1^+ + 2R_1\mathcal{R}_+^{(2)}, \qquad f_4 = \frac{i\,\mathcal{R}_-^{(2)}\,e^{2 i k_g R_2}}{k_g}.
\end{equation}
In these expressions, we have used the relations between $\mathcal{R}_\pm^{(1,2)}$ and $\,_{E,B}\mathcal{S}^{\pm}_{1,2}$, given by \ref{Rel_1}.

Moreover, the expressions for $\widetilde{\mathcal{Q}}_{1,2}(x,y,z)$ are
\begin{equation}
    \widetilde{\mathcal{Q}}_1(x,y,z) = A_x B_x^{(0)} + A_y B_y^{(0)} +\left(\frac{A_l}{2}- A_b\right) B_z^{(0)} - ik_g \phi_{20}(z),
\end{equation}
and
\begin{equation}
    \widetilde{\mathcal{Q}}_2(x,y,z) = -A_x E_x^{(0)} - A_y E_y^{(0)} -\left(\frac{A_l}{2}- A_b\right) E_z^{(0)} + ik_g \phi_{10}(z).
\end{equation}
Note that $\widetilde{\mathcal{Q}}_{1,2}(x,y,z)$ are functions only of $z$, which is the required condition.

\subsection{Remaining expressions from \ref{Section: Time-independent magnetic field}}\label{Appendix: Remaining expressions 2}
In this appendix, we will provide the remaining equations and expressions from \ref{Section: Time-independent magnetic field}.

\subsubsection{Intermediate equations and solutions}\label{Appendix: Remaining expressions 2a}
In this section, we will write down the remaining equations from the intermediate steps used to obtain the solutions in \ref{Section: Time-independent magnetic field}. 

As discussed in the main text, the second equations in \ref{Condition_1} and \ref{Condition_2} provide second-order differential equations for $f_3(y)$ and $f_2(y)$, respectively. Solving these equations yields the expressions for $f_3(y)$ and $f_2(y)$, respectively:
\begin{multline}\label{f3y}
    f_3(y) = K_{30} + K_{31}y - i\alpha k_g\left(A_b - \frac{A_l}{2}\right)y^2 - ik_g\int\mathrm{d}z\,\partial_y \phi_{20}(y,z) + ik_g e^{-2ik_gz}\int\mathrm{d}z\,e^{2ik_gz}\partial_y \phi_{20}(y,z) 
    \\
    - f_4(y)\,e^{-2ik_gz} + 2k_g^2\int\mathrm{d}y\,\phi_{20}(y,z)
\end{multline}
and
\begin{equation}\label{f2y}
    f_2(y) = K_{20} + K_{21}y + ik_g\int\mathrm{d}z\,\partial_y \phi_{10}(y,z) - ik_g e^{-2ik_gz}\int\mathrm{d}z\,e^{2ik_gz}\partial_y \phi_{10}(y,z) - f_1(y)\,e^{-2ik_gz} - 2k_g^2\int\mathrm{d}y\,\phi_{10}(y,z),
\end{equation}
where $K_{20,21,30,31}$ are constants of integration. In general, these constants $K_{20,21,30,31}$ can be functions of $z$. Since $f_2(y)$ or $f_3(y)$ is a function only of $y$, we choose these constants independent of $z$ to obtain an analytic solution.

Since $f_2(y)$ and $f_3(y)$ should be functions only of $y$, the conditions $\partial_z f_2(y) = 0$ and $\partial_z f_3(y) = 0$ yield the expressions for $f_1(y)$ and $f_4(y)$, respectively:
\begin{equation}\label{f1y}
    f_1(y) = -ik_g\left[e^{2ik_gz}\int\mathrm{d}y\,\partial_z \phi_{10}(y,z) + \int\mathrm{d}z\,e^{2ik_gz}\partial_y \phi_{10}(y,z)\right]
\end{equation}
and
\begin{equation}\label{f4y}
    f_4(y) = ik_g\left[e^{2ik_gz}\int\mathrm{d}y\,\partial_z \phi_{20}(y,z) + \int\mathrm{d}z\,e^{2ik_gz}\partial_y \phi_{20}(y,z)\right].
\end{equation}
Note that $f_4(y)$ is the same as $-f_1(y)$ with $\phi_{10}(y,z)$ replaced by $\phi_{20}(y,z)$.

Since $f_1(y)$ and $f_4(y)$ should be functions only of $y$, $\partial_z f_1(y) = 0$ and $\partial_z f_4(y) = 0$ yield integro-differential equations for $\phi_{10}(y,z)$ and $\phi_{20}(y,z)$, respectively. These integro-differential equations satisfied by $\phi_{10}(y,z)$ and $\phi_{20}(y,z)$ can be collectively written as a single equation, given by \ref{IDE} in the main text.

\subsubsection{Unknown functions}\label{Appendix: Remaining expressions 2b}
In this section, we provide the expressions for the unknown functions obtained using the formalism, which are essential for writing the components of the induced electromagnetic fields. We will also provide the expressions for $\widetilde{\mathcal{Q}}_{1,2}(x,y,z)$ introduced in the formalism.

The functions $f_{1,2,3,4}(x,y)$ are
\begin{equation}
    f_1(x,y) = -iC_{12}k_g y, \qquad f_4(x,y) = iC_{22}k_g y,
\end{equation}
\begin{equation}
    f_2(x,y) = K_{20} +(K_{21}- 2C_{11}k_g^2)y + \frac{3+2k_g^2y^2}{6}\,C_{10}\,y,
\end{equation}
and
\begin{equation}
    f_3(x,y) = K_{30} + (K_{31}+2C_{21}k_g^2)y - \frac{3+2k_g^2y^2}{6}\,C_{20}\,y - ik_g\alpha y^2\left(A_b-\frac{A_l}{2}\right).
\end{equation}
Now $\widetilde{\mathcal{Q}}_{1,2}(x,y,z)$ are
\begin{equation}
    \widetilde{\mathcal{Q}}_1(x,y,z) = \frac{2iK_{31}+\alpha A_y(i+6k_gz)}{4k_g}, \qquad \widetilde{\mathcal{Q}}_2(x,y,z) = \frac{2iK_{21}+\alpha A_x(3i+2k_gz)}{4k_g}.
\end{equation}
Note that $\widetilde{\mathcal{Q}}_1(x,y,z)$ and $\widetilde{\mathcal{Q}}_2(x,y,z)$ are functions only of $z$, which was the required condition. Now, we can write the solution.

\subsection{Remaining expressions from \ref{Section: GW passing through astrophysical magnetic fields: neutron star magnetosphere}: Stokes parameters}\label{Appendix: Remaining expressions 3}
In this appendix, we will provide the expressions for the Stokes parameters of the induced EMW signal from \ref{Section: GW passing through astrophysical magnetic fields: neutron star magnetosphere}. The Stokes parameters of the induced EMW emanating through a particular point from the stellar magnetosphere (identified by $\theta$, see \ref{Figure:Schematic description of a star's magnetosphere}) are
\begin{equation}
    I = \frac{k_g^2 (B_y^{(0)})^2 R_{LC}^2\cos^2\theta}{2}\left(|A_+|^2 + |A_\times|^2 + \frac{|A_l|^2}{4} +  |A_+||A_l|\cos{(\delta_+-\delta_l)}\right),
\end{equation}
\begin{equation}
    Q = \frac{k_g^2 (B_y^{(0)})^2 R_{LC}^2\cos^2\theta}{2}\left(|A_+|^2 - |A_\times|^2 + \frac{|A_l|^2}{4} +  |A_+||A_l|\cos{(\delta_+-\delta_l)}\right),
\end{equation}
\begin{equation}
    U = k_g^2 (B_y^{(0)})^2 R_{LC}^2\cos^2\theta \left[|A_+||A_\times| \cos(\delta_+ - \delta_\times) +\frac{|A_l||A_\times|}{2} \cos(\delta_l - \delta_\times) \right],
\end{equation}
and
\begin{equation}
    V = -k_g^2 (B_y^{(0)})^2 R_{LC}^2\cos^2\theta \left[|A_+||A_\times| \sin(\delta_+ - \delta_\times) +\frac{|A_l||A_\times|}{2} \sin(\delta_l - \delta_\times) \right].
\end{equation}
For the angle-averaged Stokes' parameters, $\cos^2\theta$ will be replaced by $1/3$.

We can also calculate the polarization angle as
\begin{equation}
    \Phi \equiv \frac{1}{2}\tan^{-1}\left(\frac{U}{Q}\right)= \frac{1}{2}\tan^{-1}\left(\frac{8|A_+||A_\times| \cos(\delta_+ - \delta_\times) +4|A_l||A_\times| \cos(\delta_l - \delta_\times)}{4|A_+|^2 - 4|A_\times|^2 + |A_l|^2 +  4|A_+||A_l|\cos{(\delta_+-\delta_l)}}\right).
\end{equation}
In general relativity, when we consider only two modes ($+$ and $\times$), these expressions will be the same, with $|A_l|$ and $\delta_l$ being zero.
\twocolumngrid
\bibliography{references}

\end{document}